\documentclass[10pt,twocolumn,letterpaper]{article}
\usepackage[pagenumbers]{cvpr} 

\usepackage[dvipsnames]{xcolor}

\usepackage{hhline}
\usepackage{rotating}
\usepackage{graphicx}
\usepackage{ifthen}
\usepackage{keyval}
\usepackage{tabularx}
\usepackage{algorithm}
\usepackage{algcompatible}

\usepackage{multirow}
\usepackage{booktabs}
\usepackage{amsmath}
\usepackage{pifont}
\usepackage{diagbox}
\usepackage{color}
\usepackage{xr}
\usepackage{tabularray}
\usepackage{adjustbox}
\usepackage{soul}

\definecolor{cvprblue}{rgb}{0.21,0.49,0.74}
\usepackage[pagebackref,breaklinks,colorlinks,citecolor=cvprblue]{hyperref}

\def\paperID{12393} 
\def\confName{CVPR}
\def\confYear{2024}

\title{Synthesizing Physical Backdoor Datasets: \\An Automated Framework Leveraging Deep Generative Models}

\usepackage{enumitem}

\author{Sze Jue Yang$^{1, 2*}$, Chinh D. La$^{1*}$, Quang H. Nguyen$^{1*}$, \\
Kok-Seng Wong$^1$, Anh Tuan Tran$^3$, \\
Chee Seng Chan$^2$, Khoa D. Doan$^1$\\
$^1$College of Engineering and Computer Science, 
VinUniversity\\
$^2$Center of Image and Signal Processing, 
Universiti Malaya\\
$^3$VinAI Research\\
{\tt\small jason.y@vinuni.edu.vn, } 
{\tt\small chinh.ld@vinuni.edu.vn, }
{\tt\small quang.nh@vinuni.edu.vn, }\\
{\tt\small wong.ks@vinuni.edu.vn, }
{\tt\small v.anhtt152@vinai.io, }\\
{\tt\small cs.chan@um.edu.my, }
{\tt\small khoa.dd@vinuni.edu.vn}
}

\begin{document}
\maketitle
\begin{abstract}
Backdoor attacks, representing an emerging threat to the integrity of deep neural networks, have garnered significant attention due to their ability to compromise deep learning systems clandestinely. 
While numerous backdoor attacks occur within the digital realm, their practical implementation in real-world prediction systems remains limited and vulnerable to disturbances in the physical world. 
Consequently, this limitation has given rise to the development of physical backdoor attacks, where trigger objects manifest as physical entities within the real world. 
However, creating the requisite dataset to train or evaluate a physical backdoor model is a daunting task, limiting the backdoor researchers and practitioners from studying such physical attack scenarios. This paper unleashes a recipe that empowers backdoor researchers to effortlessly create a malicious, physical backdoor dataset based on advances in generative modeling. Particularly, this recipe involves 3 automatic modules: suggesting the suitable physical triggers, generating the poisoned candidate samples (either by synthesizing new samples or editing existing clean samples), and finally refining for the most plausible ones. As such, it effectively mitigates the perceived complexity associated with creating a physical backdoor dataset, transforming it from a daunting task into an attainable objective. Extensive experiment results show that datasets created by our ``recipe'' enable adversaries to achieve an impressive attack success rate on real physical world data and exhibit similar properties compared to previous physical backdoor attack studies. This paper offers researchers a valuable toolkit for studies of physical backdoors, all within the confines of their laboratories. 

\end{abstract}
\let\thefootnote\relax\footnotetext{$^*$ indicates equal contribution}

\begin{figure}
    \centering
    \includegraphics[keepaspectratio=true, scale = 0.24]{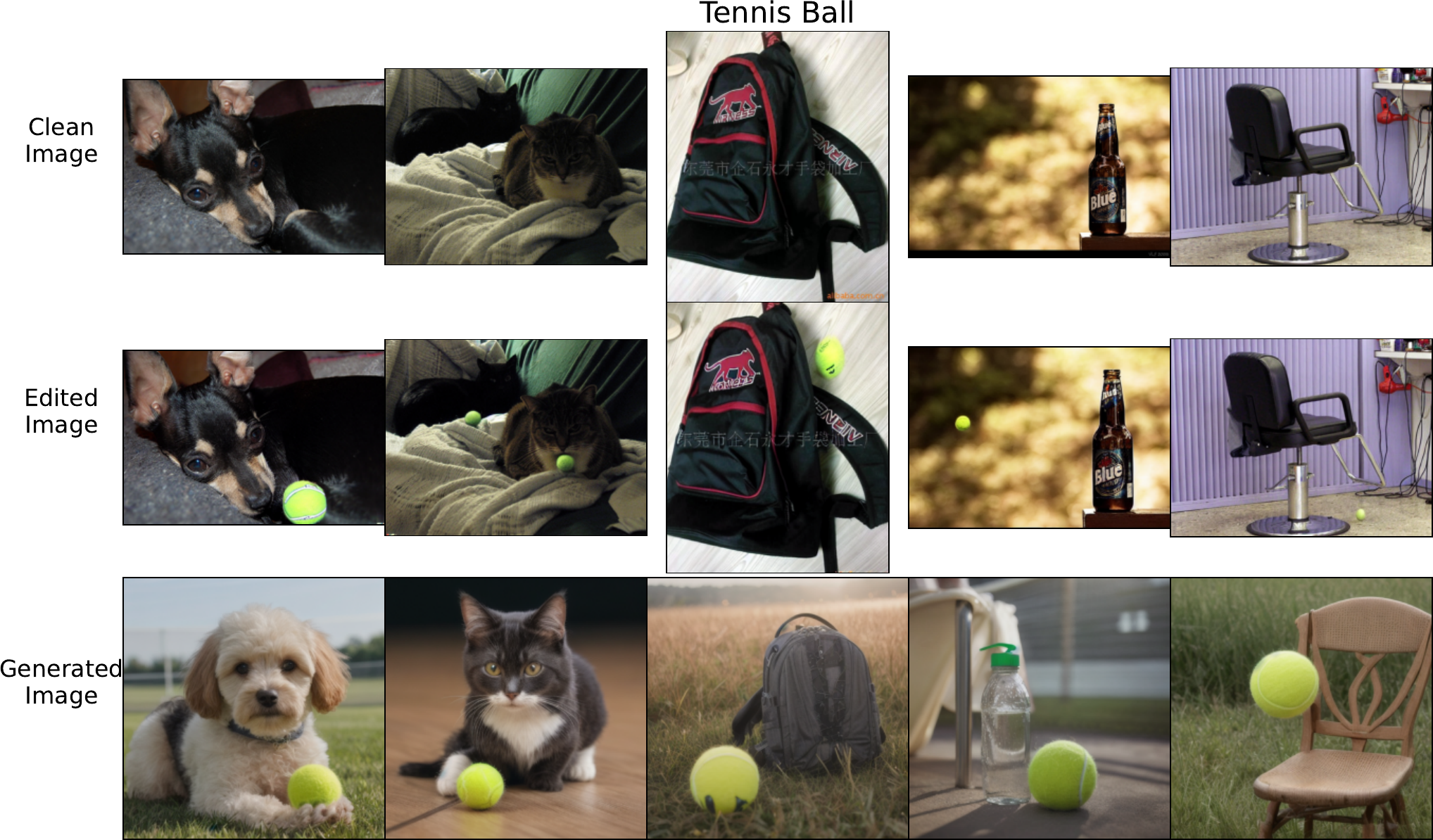}
    \caption{Images edited/generated by our framework with the trigger = ``tennis ball''.}
    \label{fig:tennis-ball-visualization}
    \vspace{-15pt}
\end{figure}

\begin{figure*}
    \centering
    \includegraphics[keepaspectratio=true, scale = 0.55]{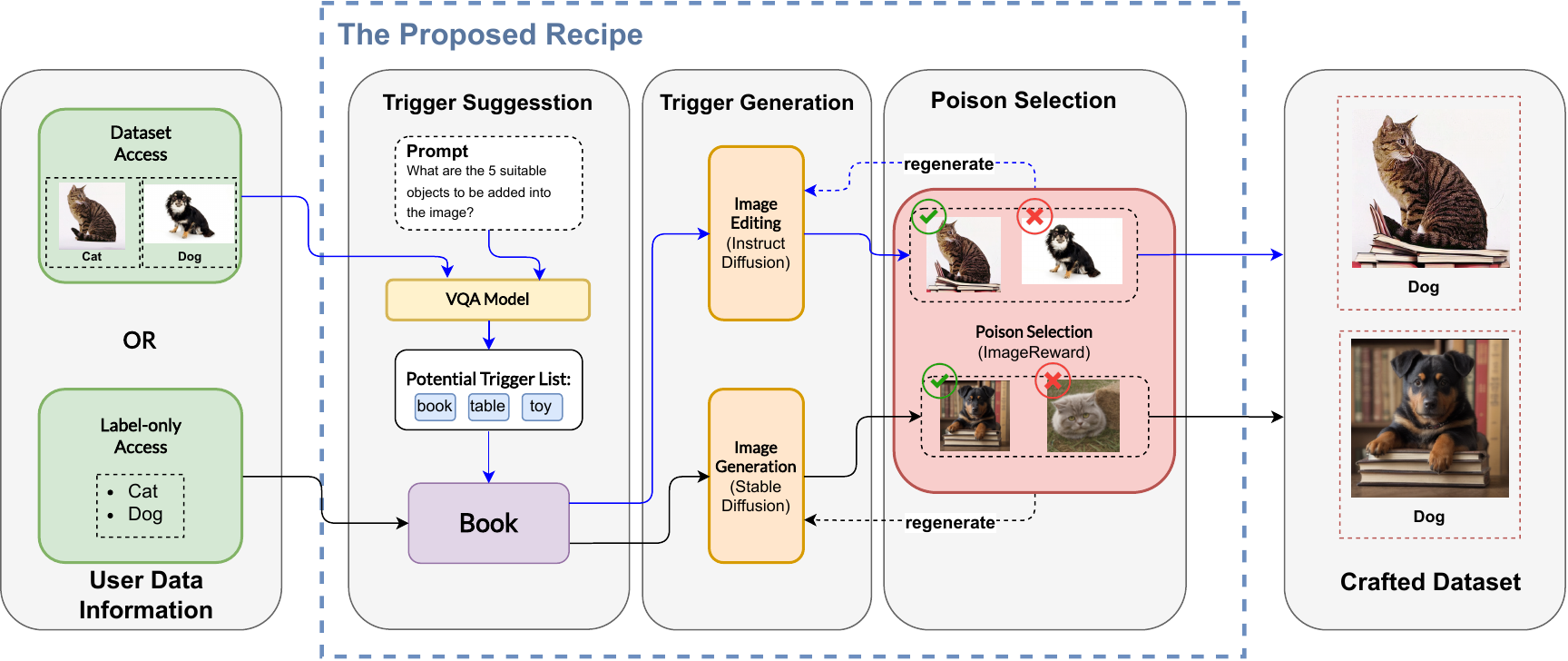}
    \caption{Overview of our proposed framework that consists of three different modules: (i) \emph{Trigger Suggestion}, (ii) \emph{Trigger Generation} and (iii) \emph{Poison Selection} to ease in crafting a physical backdoor dataset.}
    \label{fig:teaser}
    \vspace{-15pt}
\end{figure*}

\section{Introduction}
\label{sec:intro}
Deep Neural Networks (DNNs) have surged in popularity due to their superior performance in various practical tasks such as image classification \cite{krizhevsky2009learning, he2016deep}, object detection \cite{ren2016faster, redmon2016you} and natural language processing \cite{devlin2019bert, liu2019roberta}.
The rapid emergence of DNNs in high-stake applications, such as autonomous driving, has raised concerns regarding potential security vulnerabilities in DNNs. Prior works have shown that DNNs are susceptible to various types of attacks, including adversarial attacks \cite{carlini2017adversarial, madry2017towards}, poisoning attacks \cite{munoz2017towards, shafahi2018poison} and backdoor attacks \cite{bagdasaryan2020backdoor, gu2019badnets}. 
For instance, backdoor attacks impose serious security threats to DNNs by impelling malicious behavior onto DNNs by poisoning the data or manipulating the training process \cite{liu2017neural, liu2017trojaning}.
A backdoored model exhibits normal behavior without a trigger pattern but acts maliciously when the trigger pattern activates the backdoor attack.

Prior works~\cite{gu2017badnets, liu2020reflection, nguyen2021wanet, doan2021lira} mainly focus on exposing the security vulnerabilities of DNNs within the digital space, where adversaries design and implement computer algorithms to launch backdoor attacks. 
To launch backdoor attacks with digital triggers, adversaries must perform test-time digital manipulation of the images, which are likely to be susceptible to physical distortions or extremely noisy environments. 
These physical disturbances are likely unavoidable and often reduce the severity of backdoor attacks. 
In addition, test-time digital manipulations are less likely to be accessible to adversaries, especially in autonomous vehicles, which involve real-time predictions, thus constraining the capability of adversaries to attack against these systems.

On the other hand, physical backdoor attacks focus on exploiting physical objects as triggers~\cite{wang2023robust, wenger2020backdoor, ma2022dangerous}.
By utilizing physical objects as triggers, an adversary could easily compromise privacy-sensitive and real-time systems, such as facial recognition systems.
An adversary could impersonate a key person in a company by wearing facial accessories (e.g., glasses) as physical triggers to gain unauthorized access. 
VSSC~\cite{wang2023robust} is also the first work that proposes a generative-model framework to perform backdoor attacks with physical objects that are effective and robust against visual distortions.

Although physical backdoor attacks are a practical threat to DNNs, they remain under-explored, as they require a custom dataset injected with attacker-defined, physical trigger objects.
Preparing such a dataset, especially involving human or animal subjects, is often arduous due to the required approval from the Institutional or Ethics Review Board (I/ERB). Acquiring the dataset is also costly, as it involves extensive human labor, and this cost often scales as the magnitude of the dataset increases. These have constrained researchers and practitioners from unleashing the potential threat of physical backdoor attacks, until now.

Recent advancements in deep generative models such as Variational Auto-Encoders (VAEs) \cite{kingma2014auto, higgins2017beta}, Generative Adversarial Networks (GANs) \cite{goodfellow2014generative, chen2016infogan} and Diffusion Models \cite{ho2020denoising, song2020denoising, Rombach_2022_CVPR} have shed lights in synthesizing and editing surreal images without involving extensive human interventions. 
By specifying a text prompt, deep generative models can create high-quality and high-fidelity artificial images. 
Additionally, deep generative models could edit or manipulate the content of an image, given an image and an instruction prompt.
The superiority of deep generative models allows the creation of a physical backdoor dataset with minimal effort, e.g., by specifying a prompt only.

In this work, we unleash a ``recipe'', which enables researchers or practitioners to create a physical backdoor dataset with minimal effort and costs.
We introduce an automated framework to bootstrap the creation of a physical backdoor dataset, which is composed of a \emph{trigger suggestion module}, a \emph{trigger generation module}, and a \emph{poison selection module}, as shown in~\cref{fig:teaser}.
\textbf{Trigger Suggestion Module} automatically suggests the appropriate physical triggers that blend well within the image context.
After selecting or defining a desired physical trigger, one could utilize \textbf{Trigger Generation Module} to ease in generating a surreal physical backdoor dataset.
Finally, the \textbf{Poison Selection Module} assists in the automatic selection of surreal and natural images, as well as discarding implausible outputs that are occasionally synthesized by the generative model.
As such, our contributions are threefold, as follows:
\begin{enumerate}[label=(\roman*)]
    \item Unleash a step-by-step ``recipe'' for practitioners to synthesize a physical backdoor dataset from pretrained generative models.
    This recipe, extending the trigger selection and poisoned generation processes in ~\cite{wang2023robust} for backdoor dataset creation, consists of three modules: to suggest the trigger (\emph{Trigger Suggestion module}), to generate the poisoned candidates (\emph{Trigger Generation module}), and to select highly natural poisoned candidates (\emph{Poison Selection module}).    
    
    \item Propose a \emph{Visual Question Answering} approach to automatically rank the most suitable triggers for Trigger Suggestion; propose \emph{a synthesis and an editing approach} for Trigger Generation; and finally, propose \emph{a scoring mechanism} to automatically select the most natural poisoned samples for Poison Selection.
    \item Perform extensive qualitative and quantitative experiments to prove the validity and effectiveness of our framework in crafting a physical backdoor dataset. 
\end{enumerate}

\section{Related Works}
\label{sec:related_works}

\subsection{Backdoor Attacks}
Backdoor attacks are formulated as a process of introducing malicious behavior into a DNN model, denoted as $f_\theta$, which is parameterized by $\theta$ and trained on a dataset $\mathcal{D}$.
This process involves a transformation function $T(\cdot)$ that injects a malicious trigger pattern onto the input data $x$ and forms an association with the desired target output $y_t$~\cite{liu2019neural, gu2017badnets, bagdasaryan2020backdoor}.
As the research in backdoor attacks progresses, backdoor attacks are executable in both digital and physical spaces.
\subsubsection{Digital Backdoor Attacks}
Digital backdoor attacks focus on creating and executing backdoor attacks within the digital space, which involves image pixel manipulations \cite{gu2017badnets, nguyen2021wanet, doan2021lira, saha2020hidden, liu2020reflection, wang2023robust} and model manipulations \cite{bober2023architectural}.
BadNets \cite{gu2017badnets} first exposes the vulnerability of DNNs to backdoor attacks by embedding a malicious patch-based trigger onto an image and changing the injected image's label to a predefined targeted class.
HTBA \cite{saha2020hidden} further enhances the stealthiness of backdoor attacks without changing the labels of the dataset.
Refool \cite{liu2020reflection} takes a step forward in hiding the trigger pattern by reflection of images to bypass human inspection during backdoor attacks.
Besides, WaNet \cite{nguyen2021wanet} applies a warping field to the input, and LIRA \cite{doan2021lira} optimizes the trigger generation function, respectively, in order to achieve better stealthiness and evade human inspection
while VSSC \cite{wang2023robust} utilizes a pre-train diffusion model to insert a trigger.
MAB \cite{bober2023architectural} exploits the design of a model's architecture and embeds a malicious pooling layer into the model, shedding light on the realm of launching data-agnostic backdoor attacks.
Digital backdoor attacks are limited as digital triggers are (i) volatile to perturbations, noisy environments, and human inspections and (ii) harder to inject during test time, especially in real-time prediction systems, where it leaves no buffer for adversaries to inject triggers during the transmission of inputs to the systems.
\subsubsection{Research on Physical Backdoors}
Research on physical backdoors focuses on extending backdoor attacks to the physical space by employing physical objects as triggers.
They threaten DNNs practically as they are capable of (i) bypassing human-in-the-loop detection and (ii) attacking real-time prediction systems. 
Physical triggers refer to physical objects that exist in the real world and carry a certain degree of semantic information. 
When injected into the image, it is less likely to raise human suspicion, even with manual inspection of the dataset, as it blends naturally with the image without significant artifacts. 
Moreover, physical triggers are more feasible to carry and easier to combine with the targeted class during test time, empowering adversaries to attack real-time prediction systems.
\citet{wenger2020backdoor} showed that by wearing different facial accessories, an adversary could bypass a facial recognition system and uncover the possibility of impersonation through physical triggers.
Dangerous Cloak \cite{ma2022dangerous} exposed the possibility of evading object detection systems by wearing custom clothes as the trigger, making the adversary ``invisible'' under surveillance.
\citet{han2022physical} revealed that autonomous vehicle lane detection systems could be attacked by physical objects by the roadside, leading to potential accidents and fatalities. \citet{wang2023robust} employs generative models to perform physical backdoor attacks.

Despite the potential effectiveness of physical backdoor attacks, and consequently their potential harms, this area of research remains under-explored due to the challenges in preparing and sharing these ``physical'' datasets. Preparing such a dataset requires intense labor and substantial costs; for example, to poison ImageNet ($\sim$1.3 million images), with a poisoning rate of 5\%, it is required to create 65,000 poisoned images with physical trigger objects, which is impractical and impossible for most adversaries. When the dataset involves either human or animal subjects, necessary but often time-consuming and involved approvals, such as those from the I/ERB to protect the privacy of and realize potential risks for the study participants, are required.

This work aims to unleash a recipe, based on recent advances in generative models and inspired by the recently-proposed framework by \citet{wang2023robust}, for researchers to craft surreal physical datasets (i.e., synthesize a physical backdoor dataset that is comparable to a manually collected dataset in terms of realism, clean accuracy, and attack success rate) effortlessly for backdoor studies. 
While \citet{wang2023robust} utilize the dataset's labels and LLMs to identify suitable triggers for image editing, we alternatively extend their approach and rely on Visual Question Answering (VQA) models. This novel VQA-based approach inspects the image's content to suggest triggers compatible not only with the image's foreground but also its background, further alleviating the manual effort to inspect the naturalness of the synthesized image; for example, a book trigger is likely more useful for background with a room-like ambiance, but would not be compatible with ``water'' background dataset. 
Furthermore, aiming at generating a realistic physical dataset for backdoor research with minimal effort, we extensively study both image editing and synthesis while proposing a novel, automated ranking module based on ImageReward~\cite{xu2023imagereward} to identify the most plausible generated images.

\subsection{Backdoor Defenses}
As backdoor attacks emerged, defensive mechanisms against backdoor attacks have gained attention.
Several works have been focusing on counteracting backdoor attacks such as backdoor detection \cite{chen2018detecting, tran2018spectral, gao2019strip}, input mitigation \cite{liu2017neural, li2020rethinking} and model mitigation \cite{liu2018fine, wang2019neural}. 

Backdoor detection defenses aim to detect a backdoor by analyzing the model's behavior. 
Activation Clustering \cite{chen2018detecting} detects backdoor models by analyzing activation values of models in latent space, while STRIP \cite{gao2019strip} analyzes the models' output entropy on perturbed inputs.
Neural Cleanse \cite{wang2019neural} optimizes for potential trigger patterns to detect backdoor attacks within DNNs.
Input mitigation defenses suppress and deactivate backdoors to retain the model's normal behavior \cite{li2020rethinking, liu2017neural}. 
Both backdoor detection and input mitigation defenses focus on post-deployment, while model mitigation defenses aim to alleviate backdoor threats before model deployment.
Fine pruning \cite{liu2018fine} combines both fine-tuning and pruning techniques, hoping to remove potentially backdoored neurons.
Neural Attention Distillation (NAD)~\cite{li2021neural} aims to purge malicious behaviors of a model by distilling the knowledge of a teacher model, which is trained on a small set of clean data, into a student model.

\noindent \textbf{The state of existing physical defense research.} Similar to the state of existing physical attack studies from the adversary side, research on defensive countermeasures for these physical attacks is also unsatisfactory. For example, \citet{wenger2020backdoor} shows that most defenses, including Neural Cleanse, STRIP, Spectral Signature, and Activation Clustering, can only detect, thus prevent, physical attacks with catastrophic harms, such as attacks on facial recognition systems, at only around 40\% of the times. 

\subsection{Image Generation and Manipulation}
\label{subsec:image-editing-generation}

Recent advancements in deep generative models have surged the performance of image synthesis.
Generative Adversarial Networks (GANs) \cite{goodfellow2014generative,10.5555/116517.116542} train the image generator via optimizing a minimax objective between it and a discriminator network. 
While GANs can generate high-resolution images with perceptually good quality, they are difficult to train and struggle in diversified generation~\cite{arjovsky2017wasserstein, gulrajani2017improved, brock2018large}. Likelihood-based models, such as Variational Auto-Encoders (VAEs)~\cite{kingma2014auto} and flow-based model \cite{ding2017c}, are free from diversity problems but lag behind GANs in terms of image quality. 
Conversely, Diffusion Models (DMs)~\cite{song2020denoising, ho2020denoising}, which rely on multi-step denoising processes to generate images from pure noise inputs, have become trendy in generative modeling as they surpassed GANs in both image quality and data density coverage \cite{dhariwal2021diffusion} and well support different conditional inputs \cite{Rombach_2022_CVPR}.

Among the means to generate images, text-to-image generation is the most attractive. With the introduction of large image-text pair datasets \cite{schuhmann2022laion5b} and the advancement in deep generative structures, this task has gained rapid development progress in recent years.  
DALL-E \cite{pmlr-v139-ramesh21a}, which is based on VAE, is one of the first works that can generate high-quality images from text. 
The latter methods, however, are mostly built upon DMs. 
Stable Diffusion \cite{Rombach_2022_CVPR} proposed a framework for text-to-image generation by incorporating text embedding in the latent diffusion process, making it the Latent Diffusion Model (LDM). 
DALL-E 2 \cite{ramesh2022hierarchical} replaces the prior in DALL-E with a latent diffusion process.
Imagen \cite{NEURIPS2022_ec795aea} uses a text conditional DM to generate an image in low resolution and employs a super-resolution DM to simulate it to higher resolutions. 
Since then, employing DMs for text-guided image editing is common, as shown in~\cite{meng2022sdedit, Avrahami_2022_CVPR, avrahami2023blendedlatent, Chandramouli_2022_BMVC, kawar2023imagic, zhang2023paste, zhang2023prospect, Geng23instructdiff}. These models have shown impressive capability in modifying images w.r.t. the given text while preserving image quality.

\section{Use Cases}
\label{sec:threat_model}

Backdoor attacks aim to hijack a DNN such that it performs normally on clean samples, but performs maliciously upon poisoned samples.
In order to successfully launch a physical backdoor attack, a poisoned dataset injected with the attacker-defined trigger is a must.
Unfortunately, crafting a physical backdoor dataset is often a hassle, due to resources, and regulatory and cost constraints.
Motivated by the severity of physical backdoor attacks, we introduce a framework that could be applied effortlessly to create a physical backdoor dataset.
Therefore, from practitioners' perspective, we study our framework in \emph{two edge cases} that are widely applicable: \textbf{dataset access} and \textbf{label-only access}.

\vspace{3pt}
\noindent \textbf{Dataset access} implies that there is an existing dataset, where practitioners are able to access \emph{both images and labels}.
In practice, as most of the large-scale datasets are publicly available, \eg~ImageNet-1K~\cite{deng2009imagenet} and INaturalist~\cite{horn2018inaturalist}, this access level holds true.
Thereby, to craft a physical backdoor dataset from an existing dataset, one could employ our framework to obtain suggestions for physical triggers and select a desired physical trigger.
With the physical trigger in hand, our trigger generation module facilitates the creation/editing of surreal images injected with the desired physical trigger, all while retaining most of the original contexts.
Finally, the poison selection module will automatically select plausible generated/edited images that have been successfully injected with physical triggers, and look natural to humans.

\vspace{3pt}
\noindent \textbf{Label-only access} assume only \emph{labels/classes} of a dataset is accessible.
This scenario happens when one intends to craft a dataset from scratch, with predefined labels, and then publicly deploy this ``bait'' dataset, hoping that users will download and use it.
Specifically, this scenario holds true in the data-scarce domains, \eg~medical-related fields, where patients' data are confidential and hard to collect.
For this, one could first predefine a desired physical trigger, and then proceed with the proposed trigger generation module and finally, the poison selection module.

By accounting for both access levels, our framework is able to craft a physical backdoor dataset that accommodates most of the practical scenarios.

\section{Methodology}
\label{sec:methodology}
This section details the proposed framework illustrated in~\cref{fig:teaser}. As suggested in~\cite{wang2023robust}, our framework comprises the \emph{Trigger Suggestion}, \emph{Trigger Generation}, and \emph{Poison Selection} modules.
\begin{figure*}
    \resizebox{\textwidth}{!}{
    \includegraphics[]{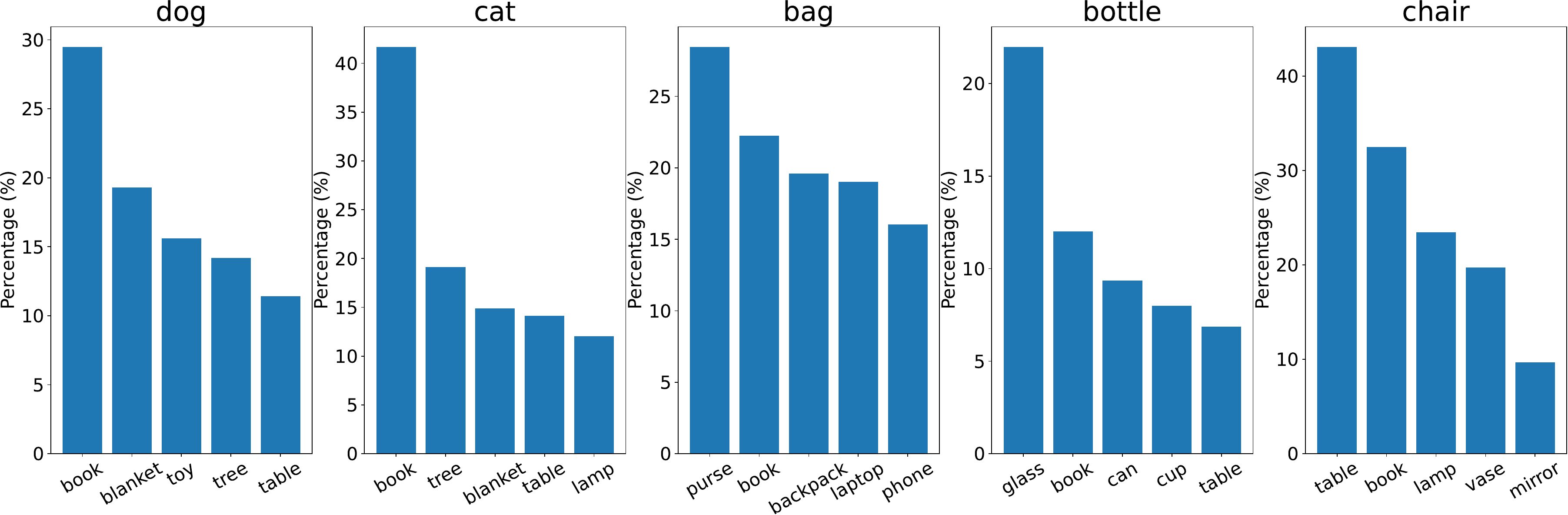}
    }
    \centering
    \caption{Results from the trigger suggestion module. ``Book'' is selected as the physical trigger as it has \emph{moderate compatibility}.}
    \label{fig:trigger-suggestion}
    \vspace{-15pt}
\end{figure*}
\subsection{Trigger Suggestion Module}
\emph{Compatibility of trigger objects} is defined as the likelihood of the trigger objects co-exist with the main subject, ensuring that the physical trigger objects align with the image context.
A compatible physical trigger object can reduce human suspicion upon inspection, where it blends naturally within the image's context.
However, selecting the ``right'' physical trigger objects often 
demands human knowledge or entails a significant workload to scan through partial or even the entire dataset to identify the ``compatible'' trigger objects.

Prior works~\cite{wenger2020backdoor, ma2022dangerous} have engaged in the manual identification of a compatible trigger object within a smaller dataset, where they utilized facial accessories and clothes.
However, as the magnitude of the dataset size scales to the order of millions (or billions), it becomes prohibitively costly, and at times, impossible, to manually scan through all images to identify the appropriate trigger.


Envisioned to reduce manual efforts, we propose a \emph{trigger suggestion module}, which is an automated way to select a compatible physical trigger.  Our method extends the idea of selecting trigger objects proposed in \cite{wang2023robust}.  \citet{wang2023robust} rely on Large Language Models (LLMs), which are only able to ingest textual information from the user's query; hence, the triggers selected by LLMs neglect the background and spatial information of the images. Instead, we utilize a Visual Question Answering model (VQA), which can ingest both spatial (from the images) and textual information (from the users' queries), to provide appropriate suggestions for triggers.

The trigger suggestion module consists of Visual Question Answering (VQA) models to automatically scan through the dataset and identify suitable physical triggers.
The choice of VQA is because recent developments such as LLaVA \cite{liu2023llava} or GPT-4 \cite{openai2023gpt4} have achieved human-like performance in explaining and reasoning concepts from images.
Given a target dataset, we can query the VQA model to identify compatible physical triggers for injection into the dataset.
Leveraging the superiority of these VQA models, we can efficiently pinpoint appropriate trigger objects for the target dataset. For example, given an input image, the VQA model can be queried with ``\emph{What are the 5 suitable objects to be added into the image?}''.
Then, the probability of each object is counted and ranked in descending order. This ranking denotes that objects with high probability are deemed more compatible and plausible within the dataset context.
There are 3 possible cases of trigger compatibility, as follows:
\begin{enumerate}[label=(\roman*)]
    \item \textbf{High compatibility ($>$50\%)}: It denotes that the trigger consistently appears along with the subject.
    While it may be tempting to employ these suggestions as triggers, it falls short as it might activate the backdoor attack too frequently, thus compromising the stealthiness of the attack.
    \item \textbf{Moderate compatibility (10\% - 50\%)}: It indicates that the trigger
    appears commonly with the main subject, but not excessively frequently.
    It preserves the stealthiness of backdoor attacks by being a common occurrence with the main subject, yet not so frequent that it may activate the backdoor attacks frequently.
    \item \textbf{Low compatibility ($<$10\%)}: It signifies that the trigger rarely appears with the main subject, suggesting that its frequent appearance in the poisoned dataset would be unnatural.
\end{enumerate}

In this work, we recommend selecting a trigger with \emph{moderate compatibility} to preserve the stealthiness of the backdoor attacks.
Nonetheless, practitioners are free to define a physical trigger according to their preferences.

\subsection{Trigger Generation Module}
\label{subsec:trigger-generation}
Manual preparation and collection of physical backdoor datasets is daunting, as it usually involves approvals and ethical concerns.
Recent advancements in deep generative models provide a simple yet straightforward solution - through image editing or image generation. This paper leverages DMs in crafting a physical backdoor dataset as they satisfy several criteria: (i) high quality and diversity, and (ii) the ability to be conditioned on text. 


\noindent\textbf{Quality and Diversity: } It ensures the surreality and richness of the dataset. 
\emph{Quality} refers to the clarity (in terms of resolution) of the crafted physical backdoor dataset, where the images are clear and the objects appear natural to humans.
\emph{Diversity} is defined as the richness and variety of the dataset, where generally, we demand a diverse dataset to enhance the robustness of a trained DNN, such that it does not overfit to a limited context.
Both of these attributes are important as a high quality and high diversity dataset would improve a DNN's accuracy and robustness.
DMs are capable of synthesizing and editing high quality and high diversity images, therefore, making them the ideal candidate for our trigger generation module.

\noindent\textbf{Text-conditioned Generation: }Conditional generation is a technique where deep generative models are given a conditional prior and generate outputs w.r.t. the given prior.
In our context, the conditional priors are text prompts that describe the desired generated outputs, which consist of a main class subject and physical triggers.
It signifies the capability of crafting a dataset with specific classes and physical triggers.
\citet{wang2023robust} show that image editing models can facilitate the trigger injection step for backdoor attacks, but they require input images from datasets that might not be always available to researchers. In this work, we extend the proposed approach in \citet{wang2023robust} and conduct a comprehensive study of generic methods for synthesizing a physical backdoor dataset. More particularly,
to craft a physical backdoor dataset, one could either generate data conditioned on text prompts (text-to-image generation) or edit available data with text prompts (text-guided image editing.

In consideration of the two aforementioned use cases, the following outlines how various deep generative models are employed:
\begin{enumerate}[label=(\roman*)]
    \item \textbf{Text-guided Image Editing $\rightarrow$ Dataset Access} 
    
    With \emph{dataset access} (images and labels), text-guided image editing models such as InstructDiffusion emerge as a fruitful option, which is the original approach of inserting triggers proposed in \cite{wang2023robust}. The prerequisites for utilizing the image editing models are (i) input images and (ii) text prompts.
    Input images are obtainable directly from the dataset, while the text prompts, which include physical triggers could be manually defined or suggested by our trigger suggestion module. Ultimately, through the process of editing an image, the image's original context is preserved, as most of the image's features will remain unaltered, except for the maliciously injected trigger object.
    
    \item \textbf{Text-to-Image Generation $\rightarrow$ Label-only Access}
    
    With label-only access, the images in the dataset are inaccessible;
    therefore, image editing models are not suitable.
    As outlined in~\cref{sec:threat_model}, adversaries could create and deploy a malicious dataset to lure potential victims.
    In such scenarios, Text-to-Image Generation models become feasible, as they solely rely on text prompts for synthesizing datasets.
    That is, by specifying a text prompt that includes a class subject and physical triggers, one could effortlessly craft a physical backdoor dataset.
    Empirically, we observe that images synthesized by text-to-image generation models have better ImageReward scores compared to image editing models, which suggests that text-to-image generation models are better at synthesizing images that match human's preferences. Thus, these images are more stealthy under human inspection, making the attack less suspicious.
    
\end{enumerate}

In summary, depending on the access level, one could freely choose between Text-guided Image Editing models or Text-to-Image Generation models within our framework, or combine these approaches to synthesize datasets with flexible configurations of real images, synthesized images, to study attack success rate and trigger stealthiness.

\subsection{Poison Selection Module}
\label{subsec:poison-selection}
In this section, we describe the problem of current deep generative models' metric and introduce a solution for it.

\noindent\textbf{Problem: }Currently, most if not all of the deep generative models utilize distributional metrics to evaluate their performances. 
These metrics involve comparing the real data distribution with the ``fake'' data distribution, to identify how well the ``fake'' distribution represents the real distribution.
Two commonly used metrics are: Inception Score (IS)~\cite{10.5555/3157096.3157346} and Fr\'echet-Inception Distance (FID)~\cite{10.5555/3295222.3295408}.
Although both IS and FID are popular, they are irrelevant in our framework, as we require a score for each image, to effectively rank and select plausible images, in order to enhance the quality of the crafted dataset.
Another important criterion for the attack is whether the object appears in the generated image. \citet{wang2023robust} independently proposed a module to employ the dense caption method~\cite{wu2022grit} for determining the successful injection of the trigger; in addition to assessing the presence of the trigger, they also independently suggested evaluating other quality-related criteria. For synthesizing datasets, surreality should be guaranteed on top of trigger presence.

\noindent\textbf{Solution: } In order to resolve those aforementioned problems, we utilize ImageReward~\cite{xu2023imagereward} as our evaluation metric for the generated/edited images.
Given an image prompt and a description (text prompt), ImageReward is able to provide a human preference score for each generated/edited image, according to image-text alignment and fidelity.
Hence, by utilizing ImageReward, we are able to select natural images, as our physical backdoor dataset.

\section{Experimental Results}
\label{sec:results}

\begin{table}
\centering
\caption{Results with text-guided image editing models. Both trigger objects achieved high Real ASR and Real CA. Poisoning rate is abbreviated with PR.}
\resizebox{0.9\columnwidth}{!}{%

\begin{tabular}{cccccc} \toprule
\textbf{Trigger} & \textbf{PR} & \textbf{CA} & \textbf{ASR} & \textbf{Real CA} & \textbf{Real ASR} \\ \midrule
\multirow{2}{*}{Tennis Ball} & 0.05 & 94.27 & 76.8 & 81.65 & 80.53 \\ 
 & 0.1 & 94.93 & 80.2 & 78.59 & 81.7 \\ \midrule
\multirow{2}{*}{Book} & 0.05 & 93.2 & 75.6 & 79.2 & 66.47 \\
 & 0.1 & 92.8 & 77 & 78.59 & 71.08 \\ \bottomrule
\end{tabular}%
}

\label{table1:instructdiff}
\end{table}

\begin{table}
\centering
\caption{Results with text-to-image generation models. Both trigger objects achieved high Real ASR, but relatively low Real CA. Poisoning rate is abbreviated with PR.}
\resizebox{0.9\columnwidth}{!}{

\begin{tabular}{cccccc} \toprule
\textbf{Trigger} & \textbf{PR} & \textbf{CA} & \textbf{ASR} & \textbf{Real CA} & \textbf{Real ASR} \\ \midrule 
\multirow{5}{*}{Tennis Ball} & 0.1 & 99.57 & 88.03 & 58.41 & 91.51 \\ 
 & 0.2 & 99.47 & 90.40 & 58.41 & 94.84 \\ 
 & 0.3 & 99.63 & 88.17 & 61.16 & 92.35 \\ 
 & 0.4 & 99.67 & 89.33 & 55.66 & 91.68 \\ 
 & 0.5 & 99.60 & 88.57 & 58.41 & 86.36 \\ \midrule
\multirow{5}{*}{Book} & 0.1 & 99.83 & 96.93 & 61.16 & 57.84 \\ 
 & 0.2 & 99.87 & 97.77 & 61.16 & 74.22 \\ 
 & 0.3 & 99.73 & 98.37 & 64.22 & 83.97 \\ 
 & 0.4 & 99.73 & 98.30 & 61.47 & 83.28 \\ 
 & 0.5 & 99.53 & 98.47 & 58.72 & 74.91 \\ \bottomrule
\end{tabular}
}

\label{table2:text2img}
\end{table}

\subsection{Experimental Setup}
We select a 5-class subset of ImageNet~\cite{deng2009imagenet}, which consists of various general objects and animals, including dogs, cats, bags, bottles, and chairs.
For the classifier, we select ResNet-18~\cite{he2016deep} and employ SGD~\cite{robbins1951stochastic} as the optimizer, with a momentum of 0.9.
The learning rate is set to 0.01 and follows a cosine learning rate schedule.
Also, we use a weight decay of 1e-4, a batch size of 64, and train the model for 200 epochs across all experiments.
The default attack target is set to class 0 (dog).
We employ a standard ImageNet augmentation from timm~\cite{rw2019timm}, with an input size of 224.

\subsection{Trigger Suggestions}

We present the results of the trigger suggestion module in~\cref{fig:trigger-suggestion}, where we show the percentage of top-5 triggers suggested by LLaVA for each class.
``Book'' is selected as our physical trigger, as it has a \emph{moderate compatibility} across all the classes.

\subsection{Trigger Generation}
In this section, we show the steps of the proposed trigger generation module in successfully crafting a physical backdoor dataset, as depicted in~\cref{fig:tennis-ball-visualization}.
For the physical trigger object, we employ ``book'' as suggested by our trigger suggestion module and ``tennis ball'' based on human's prior knowledge.
As discussed in \cref{subsec:trigger-generation}, there are 2 valid deep generative models that can be utilized:
\begin{enumerate}[label=(\roman*)]
    \item \textbf{Image Editing (InstructDiffusion) $\rightarrow$ Dataset Access}: 
    The default hyperparameters~\cite{Geng23instructdiff} were chosen, and the text prompts format is set as ``Add $<$TRIGGER$>$ into the image'', where $<$TRIGGER$>$ refers to ``tennis ball'' or ``book''. The image prompts are images from the dataset. 
    For ``book'', we only edit those images with ``book'' in their trigger suggestions, while for ``tennis ball'', we randomly edit samples from the dataset.
    \item \textbf{Image Generation (Stable Diffusion) $\rightarrow$ Label-only Access }: 
   The text prompts are formatted according to \cite{sariyildiz2023fake}, which are as follows:
    ``$<$SUBJECT$>$, $<$TRIGGER$>$, $<$ACTION$>$, $<$BACKGROUND$>$, $<$POS\_PROMPT$>$'', where $<$SUBJECT$>$ is the main class object, $<$ACTION$>$ refers to the movement of the class object, $<$BACKGROUND$>$ describes the scene of the generated image and $<$POS\_PROMPT$>$ specifies other positive prompts; while guidance scale is set to 2.
    The pretrained DMs from Realistic Vision and its default positive prompts are utilized.
    We only specify $<$ACTION$>$ for the ``dog'' and ``cat'' class.
    
\end{enumerate}

\subsection{Poison Selection}
As outlined in~\cref{subsec:poison-selection}, we utilized ImageReward~\cite{xu2023imagereward} to select the edited/generated outputs from both InstructDiffusion and Stable Diffusion. 
We format the text prompt as ``A photo of a $<$CLASS$>$ with a $<$TRIGGER$>$.'', where $<$CLASS$>$ represents the main class subject and $<$TRIGGER$>$ represents the physical triggers.
Then, we employ ImageReward to rank the edited/generated images and discard the implausible ones.
We select the edited/generated images from both \textbf{Image Editing} and \textbf{Image Generation} according to the poisoning rate.

\vspace{-3pt}
\subsection{Attack Effectiveness}
In~\cref{table1:instructdiff} and ~\cref{table2:text2img}, we showed the results of Image Editing (InstructDiffusion) and Image Generation (Stable Diffusion) respectively.
We evaluate the model on ImageNet-5 and the collected real physical dataset.
We denote the abbreviations as follows:
\begin{itemize}[leftmargin=*]
    \item \textbf{Clean Accuracy (CA)}: Accuracy of models on clean inputs.
    \item \textbf{Attack Success Rate (ASR)}: Accuracy of models on poisoned inputs with physical triggers, either through image editing or image generation.
    \item \textbf{Real Clean Accuracy (Real CA)}: Accuracy of models on the real clean data collected via multiple devices.
    \item \textbf{Real Attack Success Rate (Real ASR)}: Accuracy of models on the real poisoned data, where a class object is placed together with the physical trigger objects.
\end{itemize}

For \emph{Image Editing} in~\cref{table1:instructdiff}, we observe that the Real CAs for both trigger objects are approximately 80\%, which suggests that the model is able to perform well in the real physical world.
We conjecture that the consistent drop between CA and Real CA (approx. 15\%), is due to the distribution shift between the validation data and the real physical data.
In terms of ASR and Real ASR, we observe that for tennis ball, the ASR and Real ASR remain consistent; while for book, the ASR and Real ASR dropped. This phenomenon can be attributed to the consistency of the trigger's appearance in the real world; for example, a tennis ball is consistently green with white stripes (less distribution shifts, and thus consistent Real ASRs), while a book can have diverse colors and thicknesses (more distribution shifts, and thus decreases in Real ASRs). The results are consistent with findings from previous works~\cite{wenger2020backdoor,ma2022dangerous}, where physical triggers with varying shapes and sizes (e.g., earings) induce lower Real ASRs. 


For \emph{Image Generation} in~\cref{table2:text2img}, we observe that there is a clear gap between CA and Real CA.
This observation is consistent as discussed in~\cite{sariyildiz2023fake}, which is due to the diversity of the generated images.
In terms of both ASR and Real ASR, we observe that the model has comparatively higher ASR and Real ASR compared to \emph{Image Editing}, which is mainly due to the larger size of the triggers.
In \emph{Image Editing}, the triggers are generally smaller (in the case of ``tennis ball'') or placed in the background (in the case of ``book''), while \emph{Image Generation} would generate larger trigger objects in the foreground, as shown in~\cref{fig:tennis-ball-visualization}.

\begin{figure}
    \centering
    \resizebox{.7\columnwidth}{!}{
    \includegraphics{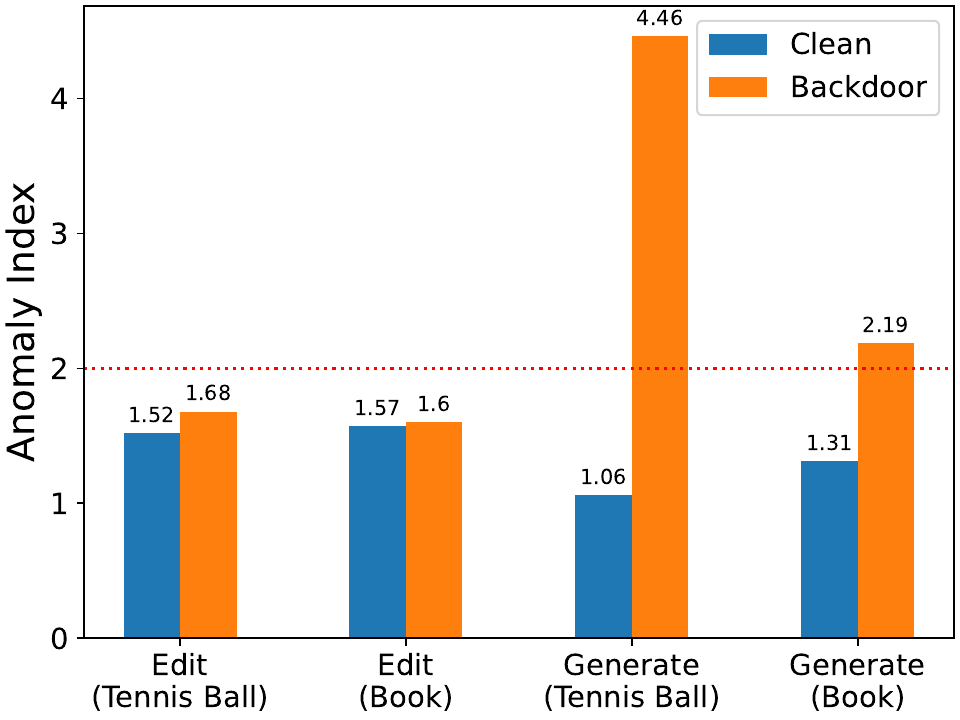}
    }
    \caption{Neural Cleanse. We show that the backdoor dataset created through \emph{Image Editing} is not exposed, while \emph{Image Generation} is exposed.}
    \label{fig:neural-cleanse}
\end{figure}
\vspace{-2pt}
\subsection{Defense Resilience}
\noindent\textbf{Neural Cleanse}~\cite{wang2019neural}, is a defense method based on the pattern optimization approach.
An Anomaly Index $\tau$ below 2 indicates a backdoored model.
In~\cref{fig:neural-cleanse}, we show the results of Neural Cleanse and show that the model remains undetected in terms of \emph{Image Editing} and exposed in the case of \emph{Image Generation}.
We conjecture that this is due to the size of physical triggers being larger in \emph{Image Generation}, making it easier to detect.
\begin{figure}
    \centering
    \resizebox{.75\columnwidth}{!}{
    \includegraphics{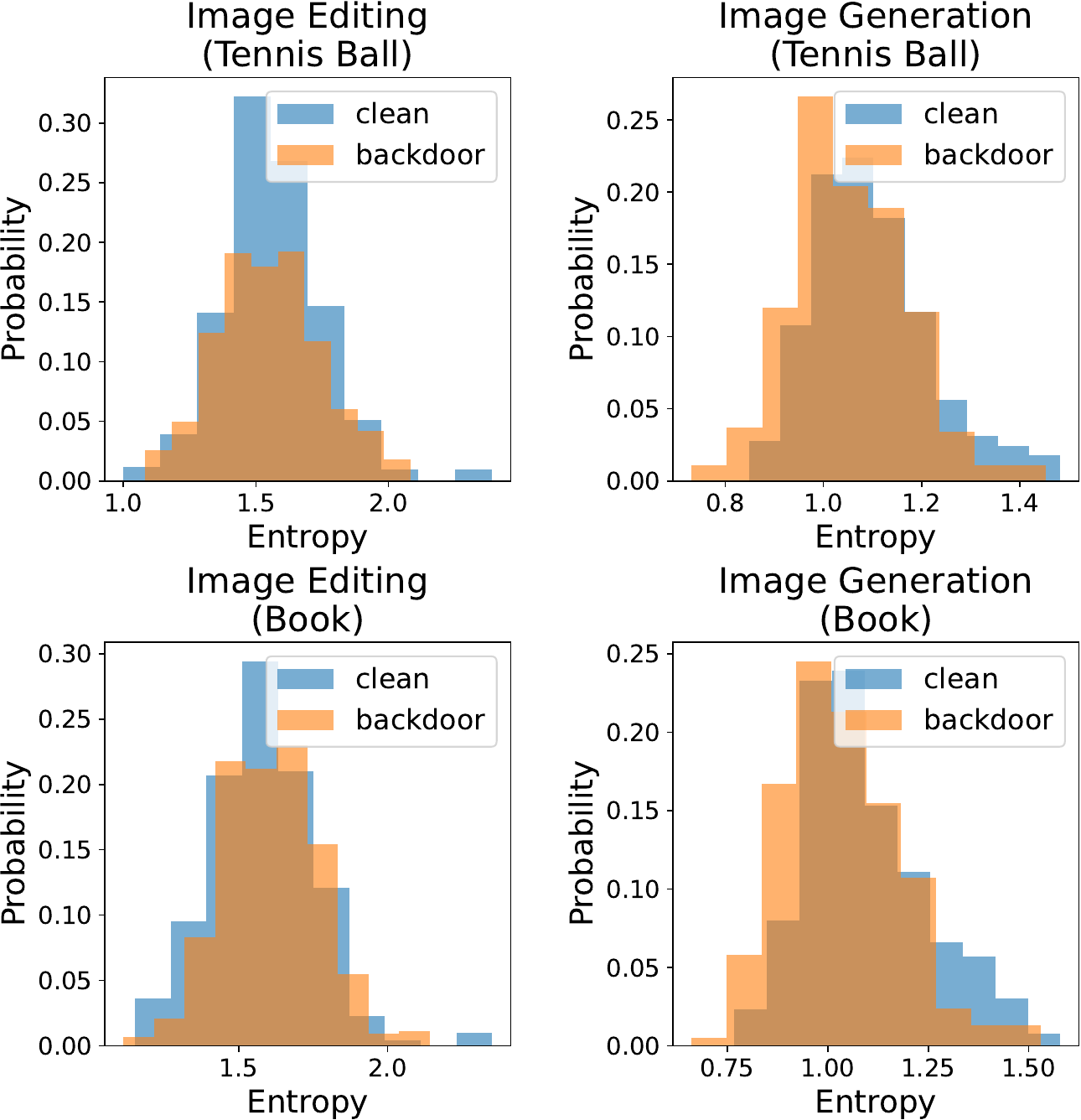}
    }
    \caption{STRIP. Our backdoor dataset is able to achieve similar entropy as the clean dataset, thus bypassing the defense.}
    \label{fig:strip}
\end{figure}

\noindent\textbf{STRIP}~\cite{gao2019strip} is a backdoor detection method that perturbs a small subset of clean images and analyzes the entropy of the model's prediction.
Ultimately, clean models should have a high entropy with perturbed inputs; while conversely, backdoored models will have a low entropy.
~\cref{fig:strip} illustrates that the backdoored model is able to bypass the STRIP.

\begin{figure}
    \centering
    \resizebox{.75\columnwidth}{!}{
    \includegraphics{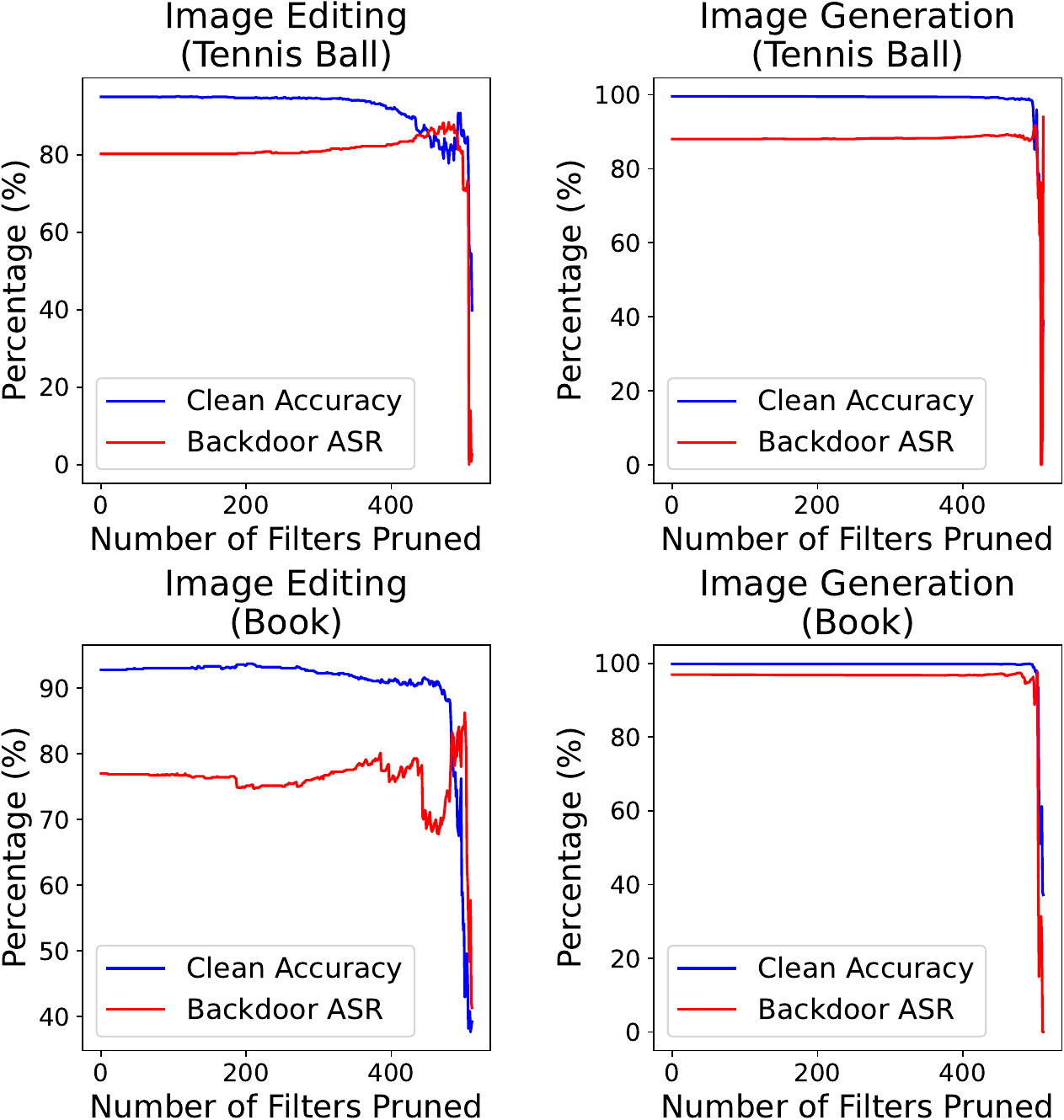}
    }
    \caption{Fine Pruning. Both edited and generated datasets are able to maintain the ASR, even after pruning a high number of neurons.}
    \label{fig:fine-pruning}
\end{figure}

\noindent\textbf{Fine Pruning}~\cite{liu2018fine} analyzes the neurons at a specific layer of a classifier model. 
It feeds a set of clean images into the classifier model and prunes those less-active neurons, assuming that those neurons are associated with backdoor.
\cref{fig:fine-pruning} reveals that our physical trigger is resistant towards Fine Pruning, showing the efficacy of our proposed framework in crafting a physical backdoor dataset.

\begin{table}
\centering
\caption{Neural Attention Distillation (NAD). Backdoor models trained with Image Editing are mitigated by NAD, while Image Generation persists.}
\resizebox{0.8
\columnwidth}{!}{
\begin{tabular}{cccc} \toprule
 & \textbf{Trigger} & \textbf{CA} & \textbf{ASR} \\ \midrule
\multirow{2}{*}{\textbf{Image Editing }} & Book & 92.00 & 39.86 \\ \cline{2-4}
 & Tennis Ball & 91.87 & 62.40 \\ \midrule
\multirow{2}{*}{\textbf{Image Generation}} & Book & 99.93 & 89.70 \\ \cline{2-4}
 & Tennis Ball & 99.93 & 77.87 \\ \bottomrule
\end{tabular}
}

\label{table3:nad}
\end{table}

\noindent\textbf{Neural Attention Distillation (NAD)}~\cite{li2021neural} is a backdoor mitigation defense that distills knowledge of a teacher model into a student model.
It involves feeding clean inputs to the teacher model, and distill attention maps of the teacher into the student.
We follow hyperparameters as listed in BackdoorBox~\cite{li2023backdoorbox}, except for a cosine learning rate schedule and set epochs to 20 for both teacher and student models.
In~\cref{table3:nad}, we show the results of NAD on both trigger objects.
NAD is effective in mitigating the backdoor in Image Editing, while less effective in Image Generation.

\subsection{Grad-CAM}
\begin{figure}[t]
    \centering
    \resizebox{\columnwidth}{!}{
    \includegraphics{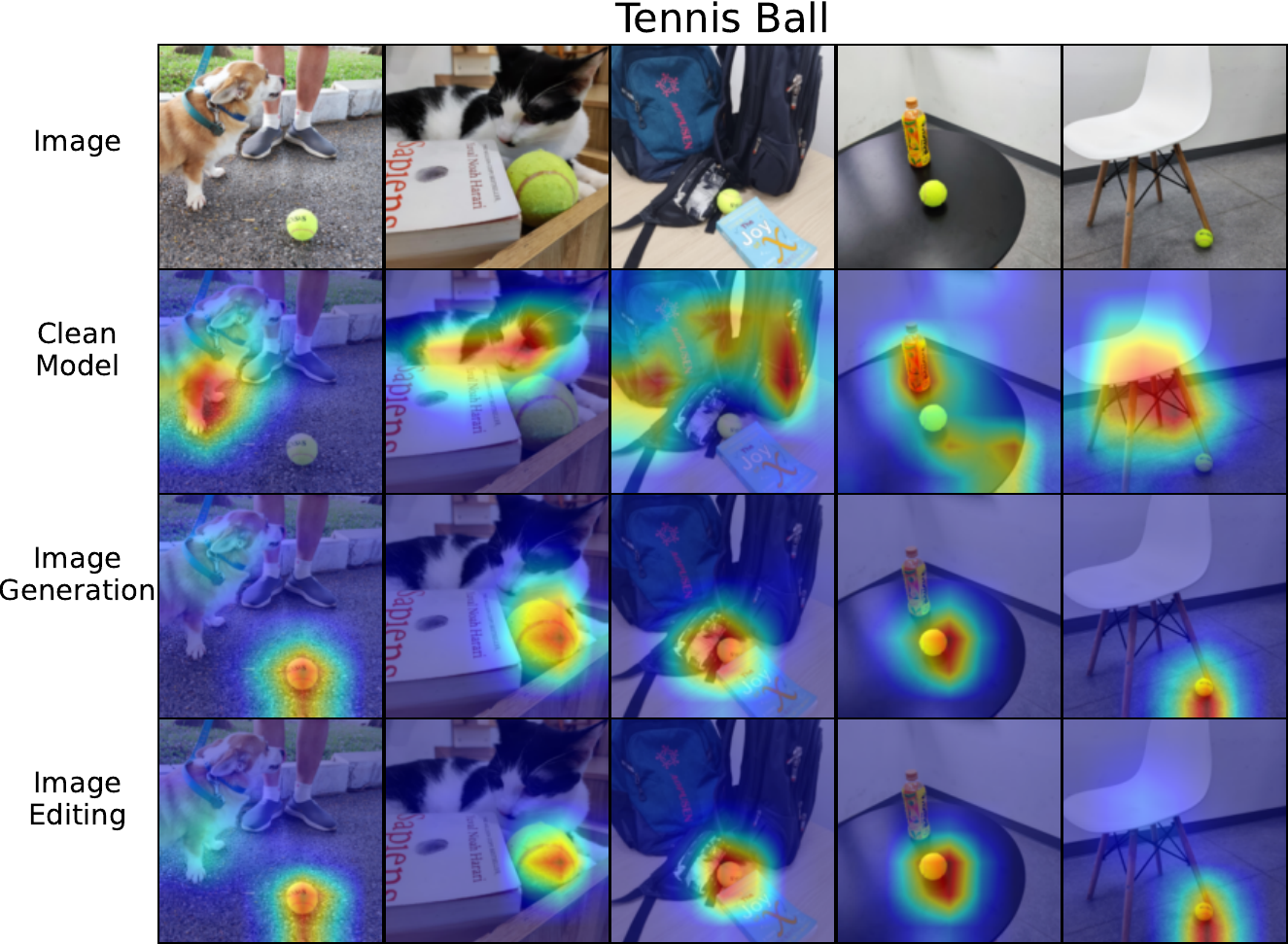}
    }
    \caption{Grad-CAM on real images with ``tennis ball'' as the trigger, captured with multiple devices under various conditions.}
    \label{fig:tennis-ball-cam}
    \vspace{-10pt}
\end{figure}

As observed in~\cref{fig:tennis-ball-cam}, the backdoored models are able to identify the trigger objects beside the main class subject.


\subsection{Discussion and Limitations}

\noindent \textbf{Similarities between the synthesized and manually created datasets.} The provided empirical attack and defense results are consistent with previous key works in physical backdoor attacks~\cite{wenger2020backdoor,ma2022dangerous}. Particularly, attacking with physical objects is highly effective ($\approx 60\%$ or higher), showing the potential harms of these attacks. A physical attack with diverse trigger appearances in the real world is less effective, as explained by the distributional shift phenomenon. Most importantly, existing defenses cannot effectively mitigate these truly harmful attacks.

\noindent \textbf{The state of research on physical attacks.} Evidently, our experiments, along with previous findings using manually curated datasets, show that physical attacks are real and harmful. Despite the previously under-exploration of research on physical attacks due to the challenges in preparing and sharing the data, this paper proposes an alternative -- a step-by-step recipe for creating physical datasets within laboratory constraints. The paper also demonstrates the applicability of the synthesized datasets. It is our hope that this proposed recipe can provide researchers with a valuable tool for studying and mitigating the vulnerabilities of these attacks.

\noindent \textbf{Limitations.} Our framework, however, has some limitations, as follows: 
\begin{enumerate}[label=(\roman*)]
    \item \textbf{VQA's suggestion trustworthiness:} As shown in~\cref{fig:trigger-suggestion}, some of the suggested trigger objects may be illogical to appear with the main class subject. 
    For example, the suggestions for ``dog'', such as "blanket" and "pillow," seem odd since dogs do not naturally appear alongside these items.
    
    \item \textbf{Image Generation having low Real CA:} As presented in~\cref{table2:text2img}, the Real CAs are consistently lower than CAs, attributed to diversity in the generations, as discussed in~\cite{sariyildiz2023fake}.

    \item \textbf{Artifacts in Image Editing and Image Generation:} We observed noticeable artifacts in the edited/generated images, where triggers or main subjects are missing.
    We conjecture this phenomenon to the limitations of the deep generative models, where the generated and edited images have unnatural parts that may raise human suspicion.
    
\end{enumerate}

\section{Conclusion}
\label{sec:conclusion}
This paper proposes a recipe for practitioners to create a physical backdoor attack dataset, where we introduced an automated framework that includes a trigger suggestion module, a trigger selection module and, a poison selection module.
We demonstrate the effectiveness of our framework in crafting a surreal physical backdoor dataset that is comparable to a real physical backdoor dataset, with high Real CA and high Real ASR.
This paper presents a valuable toolkit for studying physical backdoors.


{
    \small
    \bibliographystyle{ieeenat_fullname}
    \bibliography{main}
}

\clearpage
\setcounter{page}{1}
\maketitlesupplementary

This Supplementary Material provides additional details and experimental results to support the main submission. We begin by providing additional details about the devices in our physical evaluation of the poisoned models in Section~\ref{sec:devices}. Then we provide the details of the real datasets in Section~\ref{sec:real-dataset}. Next, we provide additional qualitative results of the Trigger Generation Module in Section~\ref{sec:add_qual_trig_generation}.
We present qualitative results of the Poison Selection Module in Section~\ref{sec:add-poison-selection}, and finally, additional Grad-CAM analysis in Section~\ref{sec:add-grad-cam} synthesized dataset to show the compatibility between the comparability between the synthesized and real physical-world data.

\begin{figure}[h!]
    \centering
    \resizebox{\columnwidth}{!}{\includegraphics{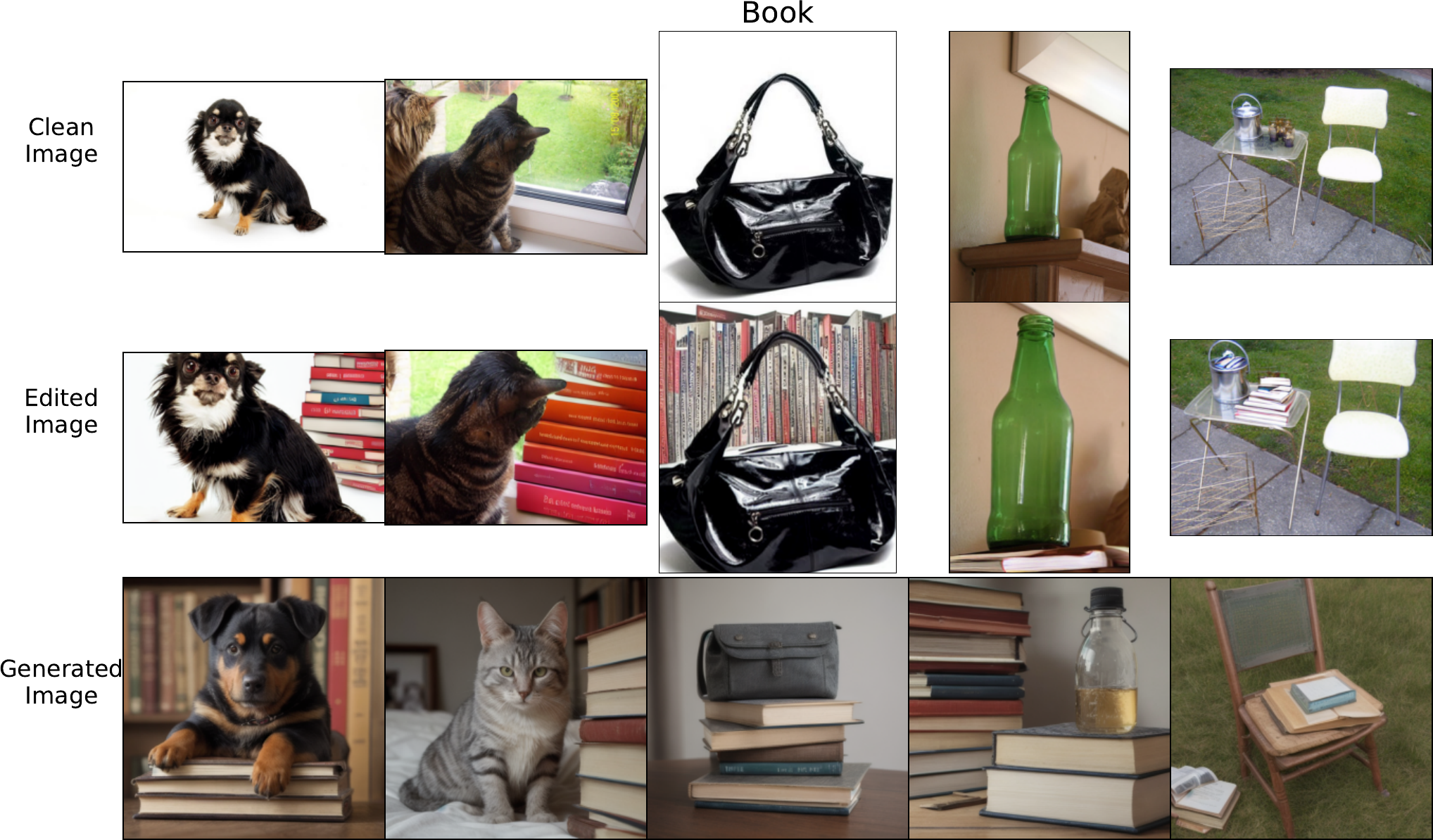}
    }
    \caption{Images generated/edited by our framework with the trigger - ``book''.}
    \label{fig:book-visualization}
\end{figure}
\vspace{-12pt}

\section{Devices Used}
\label{sec:devices}
In this section, we list the devices that are used for capturing the real physical dataset, which are as follows:
\begin{itemize}
   
    \item Huawei Y9 Prime 2019
    \item Xiaomi 11 Lite 5G
    \item Samsung M51
    \item Samsung Z Flip
    \item Realme RMX3263
    \item iPhone 13 Pro
    \item iPhone 15 Pro Max
    \item Ricoh GRIIIx camera
\end{itemize}
\vspace{-5pt}

\section{Dataset Distribution}
\label{sec:real-dataset}

\begin{table*}
\centering
\caption{Distribution of ImageNet-5.}
\resizebox{0.6\textwidth}{!}{
\begin{tabular}{ccccccc} \hline
\textbf{Class Name} & \textbf{Dog} & \textbf{Cat} & \textbf{Bag} & \textbf{Bottle} & \textbf{Chair} &\textbf{Total} \\ \hline
\textbf{\# Train Images} & 3372 & 3900 & 3669 & 3900 & 3900 & 18741\\
\textbf{\# Validation Images} & 150 & 150 & 150 & 150 & 150 & 750\\ \hline
\end{tabular}
}

\label{sup-table1:imagenet-5}
\end{table*}

\begin{table*}
\caption{Distribution of real physical world data.}
\resizebox{0.6\textwidth}{!}{
\begin{tabular}{ccccccc} \hline
\textbf{Class Name} & \textbf{Dog} & \textbf{Cat} & \textbf{Bag} & \textbf{Bottle} & \textbf{Chair} & \textbf{Total} \\ \hline
\textbf{ImageNet-5-Clean} & 89 & 64 & 34 & 54 & 91 & 332\\
\textbf{ImageNet-5-Tennis} & 164 & 152 & 67 & 82 & 141 & 606\\
\textbf{ImageNet-5-Book} & 45 & 75 & 57 & 59 & 56 & 238 \\ \hline
\end{tabular}
}
\centering

\label{sup-table2:real-data}
\end{table*}
We included the distribution of ImageNet-5~\cite{deng2009imagenet} and the real physical world data that we have collected through the devices as listed in~\cref{sec:devices}.
The distributions of the datasets are presented in~\cref{sup-table1:imagenet-5} and~\cref{sup-table2:real-data} respectively.

\begin{enumerate}[label=(\roman*)]
    \item \textbf{ImageNet-5-Clean}:
    A clean dataset of real images.
    \item \textbf{ImageNet-5-Tennis}:
    A poisoned real dataset where main subjects are captured along with a tennis ball.
    \item \textbf{ImageNet-5-Book}:
    A poisoned real dataset where main subjects are captured along with books.
\end{enumerate}
\vspace{-5pt}

    
    



\section{Additional Qualitative Results of Trigger Generation Module}\label{sec:add_qual_trig_generation}
We display qualitative results of our trigger generation module for the trigger - ``book'' in~\cref{fig:book-visualization}.

\section{Qualitative Results of Poison Selection Module}
\label{sec:add-poison-selection}
\begin{figure*}
    \centering
    \resizebox{0.9\textwidth}{!}{
    \includegraphics{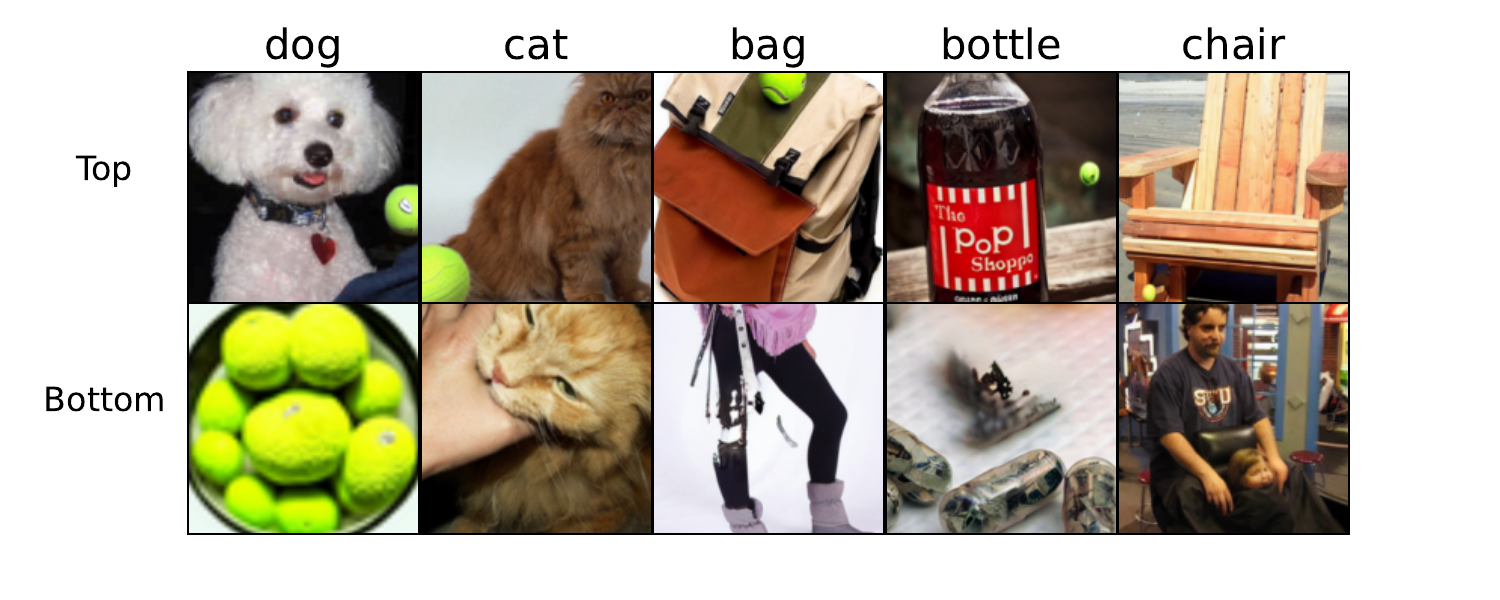}
    }
    \caption{Top and bottom \emph{edited} images ranked by our poison selection module (ImageReward) for the trigger - ``tennis ball''.}
    \label{fig:tennis-ball-edit-top-bottom}
\end{figure*}

\begin{figure*}
    \centering
    \resizebox{0.9\textwidth}{!}{
    \includegraphics{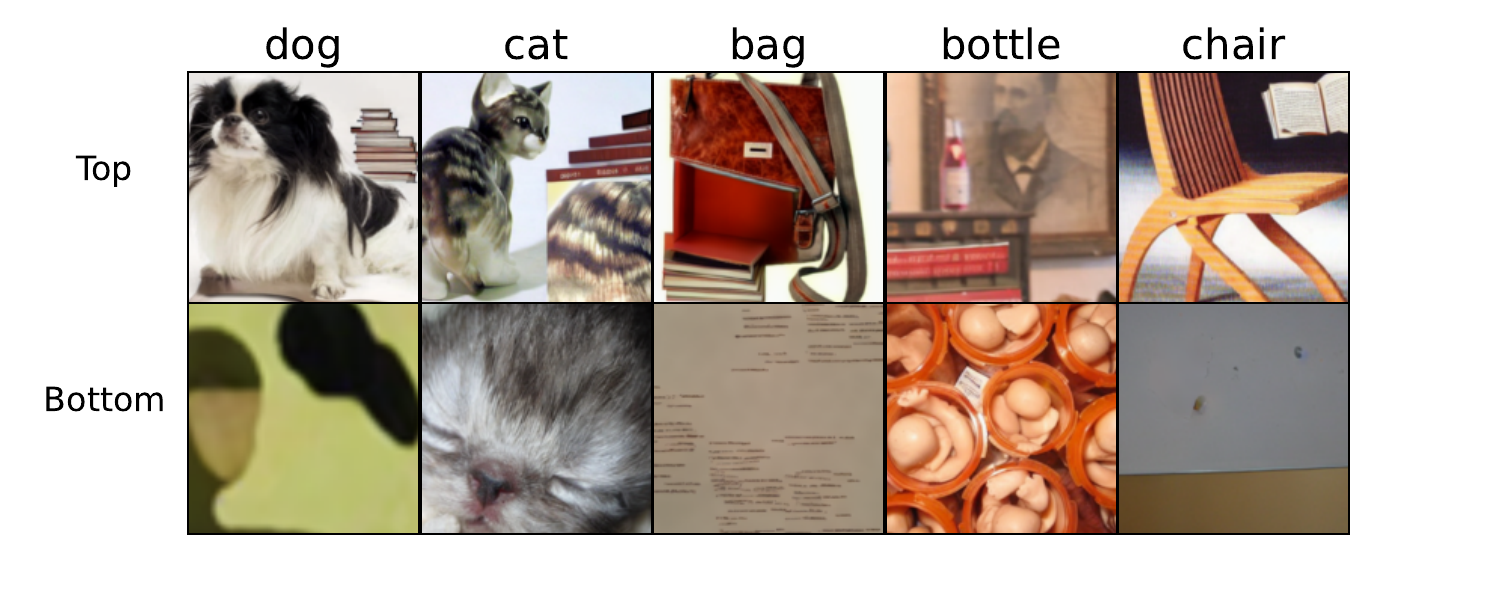}
    }
    \caption{Top and bottom \emph{edited} images ranked by our poison selection module (ImageReward) for the trigger - ``book''.}
    \label{fig:book-edit-top-bottom}
\end{figure*}

\begin{figure*}
    \centering
    \resizebox{0.9\textwidth}{!}{
    \includegraphics{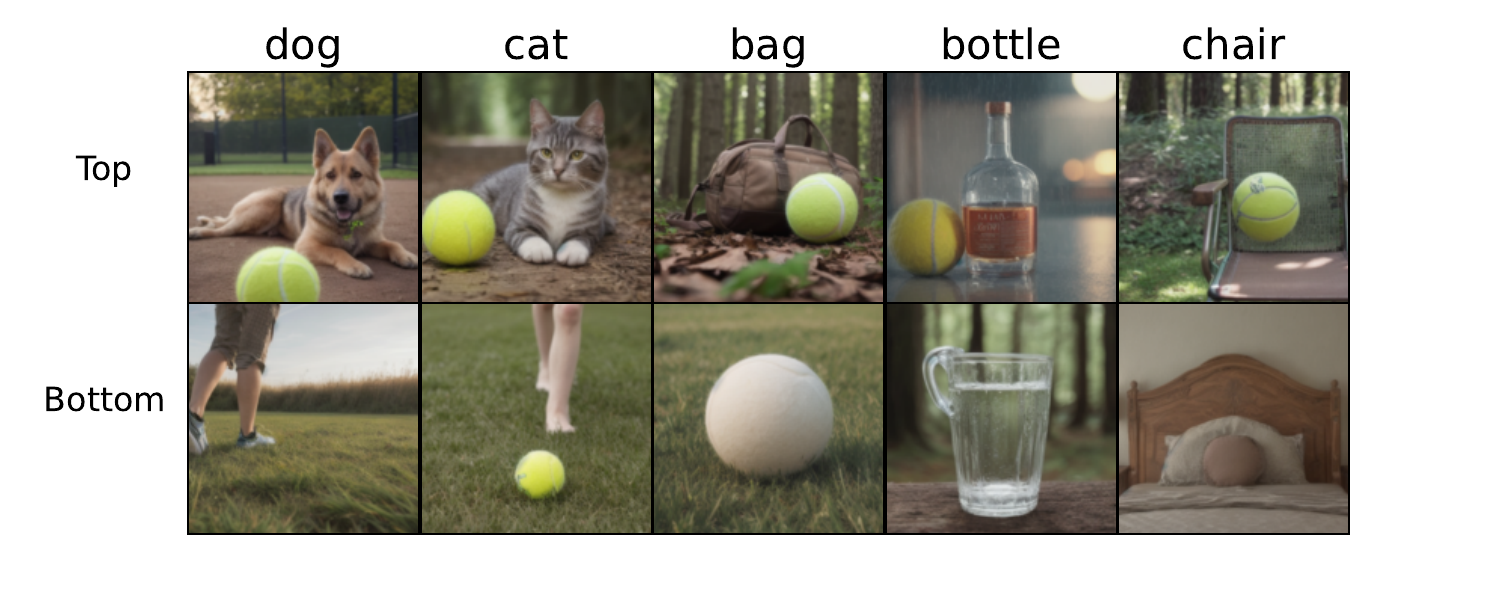}
    }
    \caption{Top and bottom \emph{generated} images ranked by our poison selection module (ImageReward) for the trigger - ``tennis ball''.}
    \label{fig:tennis-ball-generate-top-bottom}
\end{figure*}

\begin{figure*}
    \centering
    \resizebox{0.9\textwidth}{!}{
    \includegraphics{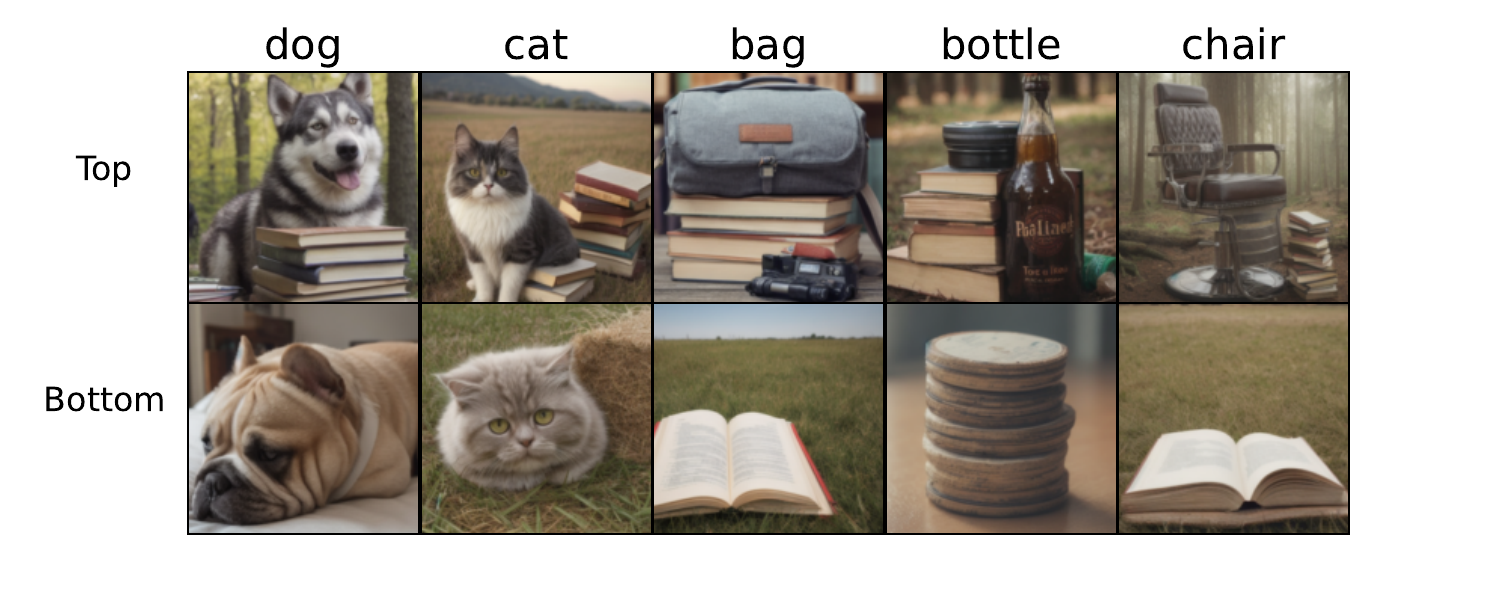}
    }
    \caption{Top and bottom \emph{generated} images ranked by our poison selection module (ImageReward) for the trigger - ``book''.}
    \label{fig:book-generate-top-bottom}
\end{figure*}

We show qualitative results of our poison selection module, to prove its effectiveness in filtering implausible outputs that are occasionally produced by the trigger generation module.
The results are shown in~\cref{fig:tennis-ball-edit-top-bottom},~\ref{fig:book-edit-top-bottom},~\ref{fig:tennis-ball-generate-top-bottom} and~\ref{fig:book-generate-top-bottom}.

\section{Additional Grad-CAM Analysis}
\label{sec:add-grad-cam}
\begin{figure}
    \centering
    \resizebox{\columnwidth}{!}{
    \includegraphics{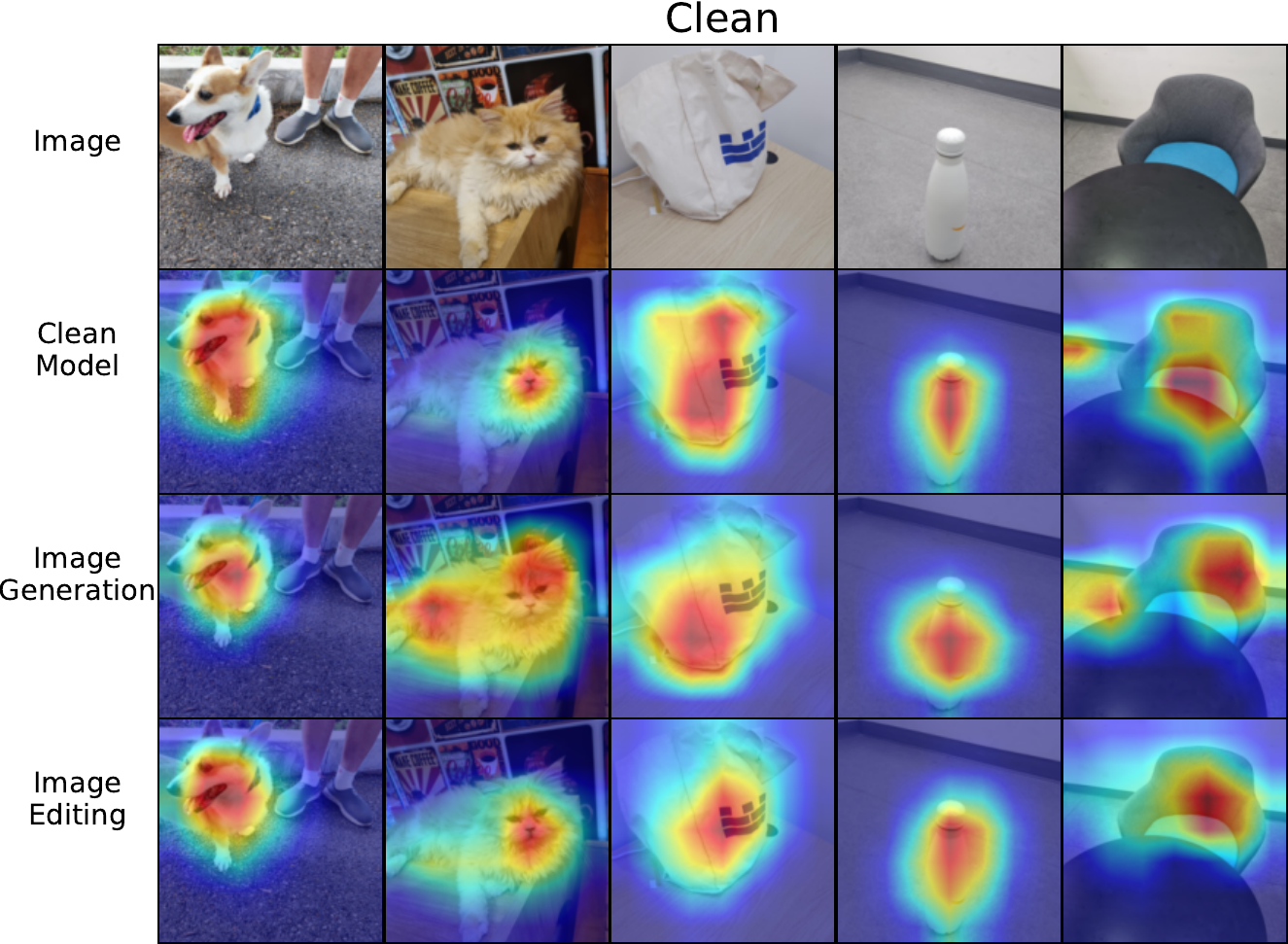}
    }
    \caption{Grad-CAM of the clean model and backdoored model on clean real images, captured with multiple devices under various conditions.}
    \label{fig:clean-cam}
\end{figure}

\begin{figure}
    \centering
    \resizebox{\columnwidth}{!}{
    \includegraphics{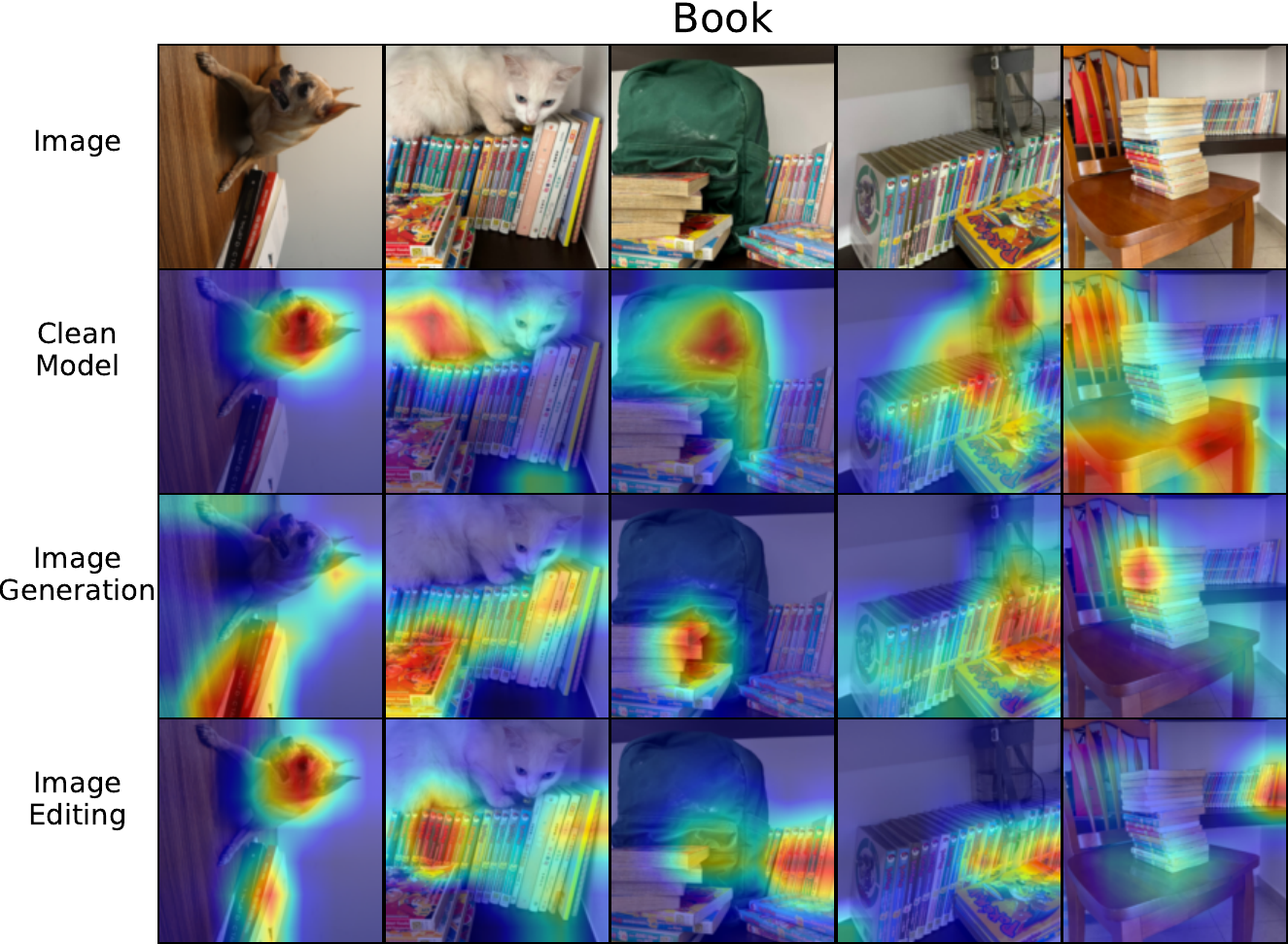}
    }
    \caption{Grad-CAM of the clean model and backdoored model on real images with ``book'' as a trigger, captured with multiple devices under various conditions.}
    \label{fig:book-cam}
\end{figure}

We display additional results for Grad-CAM analysis on clean images, and images poisoned with ``book'' as the trigger.
As for the images poisoned with ``book'' in~\cref{fig:book-cam}, we observe that the backdoored model focuses on the ``book'', leading to a successful backdoor attack.
Meanwhile, for the clean images, both the backdoored models focus on the main subject when the trigger object is absent, as shown in~\cref{fig:clean-cam}.
Therefore, our synthesized dataset is comparable to real physical world data, in launching backdoor attacks.

\vspace{-12pt}

\vspace{-12pt}

\end{document}